\documentclass[10pt,twoside]{lyxor_article}

\usepackage{amssymb}
\usepackage{amsmath}
\usepackage[T1]{fontenc}
\usepackage{lmodern}
\usepackage[latin9]{inputenc}
\usepackage{amsfonts}
\usepackage{graphicx}
\usepackage{url}
\usepackage{fancyhdr}
\usepackage[linktocpage,colorlinks=true,linkcolor=black,filecolor=black,urlcolor=black]{hyperref}

\def\tableskip{\vskip 10pt plus 2pt minus 2pt\relax}
\def\figureskip{\vskip 10pt plus 2pt minus 2pt\relax}

\newlength{\figurewidth}
\newlength{\figureheight}
\newtheorem{remark}{Remark}

\def\limfunc#1{\mathop{\rm #1}}

\newcommand{\bP}{\mathrel{\;\;}}

\newcommand{\bPP}{\mathrel{\;\;\;\;}}

\begin{document}

\title{\textbf{\color{lyxor_dark_blue}A Fast Algorithm for Computing High-dimensional Risk Parity Portfolios%
\footnote{We are grateful to Florin Spinu for providing us his code on the Newton-Nesterov algorithm
and for stimulating discussions on risk parity optimization.}}}

\author{
{\color{lyxor_dark_blue} Théophile Griveau-Billion} \\
Quantitative Research \\
Lyxor Asset Management, Paris \\
\texttt{theophile.griveau-billion@lyxor.com} \and
{\color{lyxor_dark_blue} Jean-Charles Richard} \\
Quantitative Research \\
Lyxor Asset Management, Paris \\
\texttt{jean-charles.richard@lyxor.com} \and
{\color{lyxor_dark_blue} Thierry Roncalli} \\
Quantitative Research \\
Lyxor Asset Management, Paris \\
\texttt{thierry.roncalli@lyxor.com} }

\date{\color{lyxor_light_blue}September 2013}

\maketitle

\begin{abstract}
In this paper we propose a cyclical coordinate descent (CCD) algorithm for solving
high dimensional risk parity problems. We show that this algorithm
converges and is very fast even with large covariance matrices ($n > 500$).
Comparison with existing algorithms also shows that it is one of the most
efficient algorithms.
\end{abstract}

\medskip


\section{Introduction}

In this paper, we focus on risk parity (or risk budgeting)
portfolios. The underlying idea is to do allocation by risk, not by
capital. In this case, the portfolio manager defines a set of risk
budgets and then compute the weights of the portfolio such that the
risk contributions match the risk budgets.\bigskip

From a mathematical point of view, a risk budgeting (or RB)
portfolio is defined as follows (Roncalli, 2013):
\begin{equation*}
\left\{
\begin{array}{l}
\mathcal{RC}_{i}\left( x\right) =b_{i}\mathcal{R}\left( x\right)  \\
b_{i}>0 \\
x_{i}>0 \\
\sum_{i=1}^{n}b_{i}=1 \\
\sum_{i=1}^{n}x_{i}=1
\end{array}
\right.
\end{equation*}
where $x_{i}$ is the weight of the asset $i$, $x=\left( x_{1},\ldots
,x_{n}\right) $ is the vector of portfolio weights and $b_{i}$ is
the risk budget of the asset $i$. $\mathcal{R}\left( x\right) $ is
the risk measure of the portfolio $x$ whereas
$\mathcal{RC}_{i}\left( x\right) $ is the risk contribution of the
asset $i$ for the portfolio $x$.\bigskip

A first route to solve the previous problem is to find the portfolio
$x$ such that:
\begin{equation*}
\frac{\mathcal{RC}_{i}\left( x\right)
}{b_{i}}=\frac{\mathcal{RC}_{j}\left( x\right) }{b_{j}}
\end{equation*}
This route has been explored by Maillard \textsl{et al.} (2010) in
the case
of the ERC portfolio\footnote{%
The ERC (or equal risk contribution) portfolio is a special case of
risk
budgeting portfolios when the risk budgets are the same ($b_{i}=b_{j}=\frac{1%
}{n}$).}. They propose to minimize the sum of squared differences
using the SQP algorithm. However, this algorithm is time-consuming
and does not always converge for high-dimensional problem, i.e. when
the number $n$ of assets is larger than 200.\bigskip

Another route is to consider the alternative optimization program
(Roncalli, 2013):
\begin{eqnarray}
\label{eq:rb1}
y^{\star} & = & \arg \min \mathcal{R}\left( y\right) \\
          & \mathrm{u.c.} & \left\{
\begin{array}{ll}
  \sum_{i=1}^n \ln y_i \geq c \\
  y \geq \mathbf{0}
\end{array} \right.
\nonumber
\end{eqnarray}
where $c$ is an arbitrary constant. The RB solution is then $x^\star
= y^\star/\left( \textbf{1}^\top y^\star \right)$ because of the
budget constraint $\mathbf{1}^\top x = 1$. This second formulation
has been used by Chaves \textsl{et al.} (2012) to define a Newton
algorithm. In a recent paper, Spinu (2013) improves the
convergence of the algorithm by noticing that the objective function
is self-concordant. In this case, one can use the tools developed by
Nesterov (2004). To our knowledge, the Newton algorithm was until now the best
algorithm to solve high-dimensional risk parity portfolios.\bigskip

In this paper, we present another competitive algorithm by noting
that the optimization problem (\ref{eq:rb1}) is very standard. It is
the minimization of a quadratic function with a logarithm barrier.
That's why we consider a cyclical coordinate descent algorithm used in
machine learning to solve regression problems with
non-differentiable constraints. It appears that the method is
very simple to implement and is very efficient to solve
high-dimensional risk parity portfolios ($n > 250$).

\section{Cyclical coordinate descent algorithm}

The main idea behind the cyclical coordinate descent (CCD) algorithm is to
minimize a function $f\left(x_1,\ldots,x_n\right)$ by minimizing
only one direction at each step, whereas classical descent
algorithms consider all the directions at the same time. In this
case, we find the value of $x_{i}$ which minimizes the objective
function by considering the values taken by $x_{j}$ for $j\neq i$ as
fixed. The procedure repeats for each direction until the global
minimum is reached. This method uses the same principles as
Gauss-Seidel or Jacobi algorithms for solving linear
systems.\bigskip

Convergence of coordinate descent methods requires that
$f\left(x\right)$ is strictly convex and differentiable. However,
Tseng (2001) has extended the convergence properties to a
non-differentiable class of functions:
\begin{equation}
\label{eq:fseparable} f\left(x_{1},...,x_{n}\right) =
f_{0}\left(x_{1},...,x_{n}\right) + \sum_{k=1}^{m}
f_{k}\left(x_{1},...,x_{n}\right)
\end{equation}
where $f_{0}$ is strictly convex and differentiable and the
functions $f_{k}$ are non-differentiable.\bigskip

Some properties make this algorithm very attractive. First, it is
very simple to understand and to implement. Moreover, the method is
efficient for solving large scale problems. That's why it is used in
machine learning theory for computing constrained regressions or
support vector machine problems (Friedman \textsl{et al.}, 2010). A further advantage is that the
method don't need stepsize descent tuning as opposed to gradient
based methods.

\begin{remark}
This algorithm has been already used in computing mean-variance optimization with norm constrained, because this problem
is very close to the lasso regression (Yen and Yen, 2013).
\end{remark}

\subsection{Derivation of the algorithm}

Let us derive the algorithm in the case where the risk measure is
the portfolio volatility. The Lagrangian function of the problem
(\ref{eq:rb1}) is given by:
\begin{equation}
\mathcal{L}\left( x;\lambda \right) =\arg \min \sqrt{x^{\top }\Sigma x}%
-\lambda \sum_{i=1}^{n}b_{i}\ln x_{i}
\label{eq:rb2}
\end{equation}
Without loss of generality, we can fix $\lambda = 1$. The first-order
conditions are
\begin{equation*}
\frac{\partial \,\mathcal{L}\left( x;\lambda \right) }{\partial \,x_{i}}=%
\frac{\left( \Sigma x\right) _{i}}{\sigma \left( x\right) }-\frac{b_{i}}{%
x_{i}}
\end{equation*}
At the optimum, we have $\partial _{x_{i}}\,\mathcal{L}\left(
x;\lambda \right) =0$ or:
\begin{equation*}
x_{i}\cdot \left( \Sigma x\right) _{i}-b_{i}\sigma \left( x\right)
=0
\end{equation*}
It follows that:
\begin{equation*}
x_{i}^{2}\sigma _{i}^{2}+x_{i}\sigma _{i}\sum_{j\neq i}x_{j}\rho
_{i,j}\sigma _{j}-b_{i}\sigma \left( x\right) =0
\end{equation*}
By definition of the RB portfolio we have $x_{i}>0$. We notice that
the polynomial function is convex because we have $\sigma
_{i}^{2}>0$. Since the
product of the roots is negative\footnote{%
We have $-b_{i}\sigma _{i}^{2}\sigma \left( x\right) <0$.}, we
always have two solutions with opposite signs. We deduce that the
solution is the positive root of the second degree equation:
\begin{equation}
x_{i}^{\star }=\frac{-\sigma _{i}\sum_{j\neq i}x_{j}\rho _{i,j}\sigma _{j}+%
\sqrt{\sigma _{i}^{2}\left( \sum_{j\neq i}x_{j}\rho _{i,j}\sigma
_{j}\right) ^{2}+4b_{i}\sigma _{i}^{2}\sigma \left( x\right)
}}{2\sigma _{i}^{2}} \label{eq:cwd1}
\end{equation}
If the values of $\left( x_{1},\cdots ,x_{n}\right) $ are strictly
positive, it follows that $x_{i}^{\star }$ is strictly positive. The
positivity of the solution is then achieved after each iteration if
the starting values are positive. The coordinate-wise descent
algorithm consists in iterating the equation (\ref{eq:cwd1}) until
convergence.

\begin{remark}
The convergence of the previous algorithm is obtained because
the function (\ref{eq:rb2}) verifies the technical assumptions (B1)-(C2)
necessary to apply Theorem 5.1 of Tseng (2001).
\end{remark}

\begin{remark}
We notice that the algorithm is not well-defined if some risk budgets are set to zero.
This enhances the specification of the risk budgeting problem with strictly positive values
of $b_i$.
\end{remark}

We can improve the efficiency of the algorithm because some
quantities may be easily updated. If we rewrite the equation
(\ref{eq:cwd1}) as follows:
\begin{equation*}
x_{i}^{\star }=\frac{-\left(\Sigma x\right)_{i}+x_{i}\sigma _{i}^{2}+%
\sqrt{\left( \left( \Sigma x\right) _{i}-x_{i}\sigma _{i}^{2}\right)
^{2}+4\sigma _{i}^{2}b_{i}\sigma \left( x\right) }}{2\sigma
_{i}^{2}}
\end{equation*}
we deduce that $\Sigma x$ and $\sigma \left( x\right) $ must be
computed at each iteration of the algorithm. We note $x=\left(
x_{1},\ldots ,x_{i-1},x_{i},x_{i+1},\ldots ,x_{n}\right) $ and
$\tilde{x}=\left( x_{1},\ldots ,x_{i-1},x_{i}^{\star
},x_{i+1},\ldots ,x_{n}\right) $ the vector of weights before and
after the update of the $i^{\mathrm{th}}$ weight $x_{i}$. Simple
algebra show that:
\begin{equation*}
\Sigma \tilde{x} = \Sigma x -\Sigma_{.,i} x_{i} + \Sigma_{.,i} \tilde{x}_{i}
\end{equation*}
and:
\begin{equation*}
\sigma \left( \tilde{x}\right) = \sqrt{\sigma ^{2}\left( x\right) -
2 x_i \Sigma_{i,.} x + x_{i}^{2} \sigma _{i}^{2} + 2
\tilde{x}_i \Sigma_{i,.} \tilde{x} - \tilde{x}_{i}^{2}\sigma
_{i}^{2}}
\end{equation*}
where $\Sigma_{i,.}$ and $\Sigma_{.,i}$ are the $i^{\mathrm{th}}$ row and column of $\Sigma$.
Updating $\Sigma x$ and $\sigma \left( x\right) $ is then
straightforward and reduces to the computation of two vector
products. These operations dramatically reduce the computational
time of the algorithm.

\subsection{Extension to standard deviation-based risk measure}

Roncalli (2013) considers the standard deviation-based risk measure:
\begin{eqnarray*}
\mathcal{R}\left( x\right)  &=&-\mu \left( x\right) +c\cdot \sigma \left( x\right)  \\
&=&-x^{\top }\mu +c\cdot \sqrt{x^{\top }\Sigma x}
\end{eqnarray*}
In this case, the update step of the cyclical coordinate descent
algorithm becomes:
\begin{equation*}
x_{i}^{\star }=\frac{-c\left( \sigma _{i}\sum_{j\neq i}x_{j}\rho
_{i,j}\sigma _{j}\right) +\mu _{i}\sigma \left( x\right) +\sqrt{\left(
c\left( \sigma _{i}\sum_{j\neq i}x_{j}\rho _{i,j}\sigma _{j}\right) -\mu
_{i}\sigma \left( x\right) \right) ^{2}+4cb_{i}\sigma _{i}^{2}\sigma \left(
x\right) }}{2c\sigma _{i}^{2}}
\end{equation*}

\section{Comparison of performances}

In this section, we compare the efficiency of five algorithms:
the SQP algorithm with BFGS update, the Jacobi algorithm, the Newton algorithm,
the Nesterov algorithm and the CCD algorithm%
\footnote{Details about these algorithms are provided in Appendix \ref{appendix:algorithm}}.
In order to obtain comparable results, we use the correlation matrix instead of the covariance matrix%
\footnote{Because scaling the weights by the volatilities gives the RB portfolio.}
and we assume that each algorithm is initialized with the equally-weighted portfolio.
We also use the same convergence criterion.
The algorithm is stopped when the normalized risk contributions satisfy the following inequality:
\begin{equation*}
\sup_{i}\left( \mathcal{RC}_{i}^{\star }-b_{i}\right) \leq 10^{-8}
\end{equation*}

We consider the smart beta application of Cazalet \textsl{et al.} (2013) with
the Eurostoxx 50 and S\&P 500 indexes from December 1989 to December 2012.
For each asset universe, we compute every month the ERC portfolio.
In Tables \ref{tab:sx5e3} and \ref{tab:spx3}, we report some statistics about the convergence
and the computational time%
\footnote{The numerical tests are done with the programming language Gauss 10 and an Intel T8400 3 GHz
Core 2 Duo processor and 3.5 GB RAM.} (measured in hundredths of a second).
$p_s$ indicates the convergence frequency, $\bar{T}$ is the average computational time for each iteration whereas
$T_{\max}$ is the maximum of the computational times. We notice that the SQP algorithm is the slowest algorithm and
do not not converge for all the rebalancing dates. We also observe the same convergence problem for the Jacobi algorithm.
Thus, it converges only 7 times out of ten in the case of the S\&P 500 ERC backtesting. In terms of speed,
the Jacobi method is the fastest method followed by the Newton algorithm when the number of assets is small ($n = 50$)
and by the CCD algorithm when the number of assets is large ($n = 500$).

\begin{table}[tbph]
\centering
\caption{Results with the Eurostoxx 50 ERC backtesting}
\label{tab:sx5e3}
\tableskip
\begin{tabular}{|c|ccccc|}
\hline
          Statistics &  SQP         &  Jacobi      &  Newton       &  Nesterov     & CCD           \\
\hline
  $p_s$              &     $94.96$ &     $92.02$ &     $100.00$ &     $100.00$ &     $100.00$ \\
  $\bar{T}$          & ${\bP}2.00$ & ${\bP}0.04$ & ${\bPP}0.04$ & ${\bPP}0.05$ & ${\bPP}0.12$ \\
  $T_{\max}$         & ${\bP}4.70$ & ${\bP}1.60$ & ${\bPP}1.60$ & ${\bPP}1.60$ & ${\bPP}1.60$ \\
\hline
\end{tabular}
\tableskip
\centering
\caption{Results with the S\&P 500 ERC backtesting}
\label{tab:spx3}
\tableskip
\begin{tabular}{|c|cccc|}
\hline
          Statistics &  Jacobi      &  Newton       &  Nesterov     & CCD            \\
\hline
  $p_s$              &     $69.14$ &     $100.00$ &     $100.00$ &      $100.00$ \\
  $\bar{T}$          & ${\bP}1.15$ & ${\bP}17.90$ & ${\bP}38.78$ &  ${\bPP}4.33$ \\
  $T_{\max}$         &     $18.80$ & ${\bP}21.90$ & ${\bP}54.70$ &  ${\bPP}9.40$ \\
\hline
\end{tabular}
\end{table}

\begin{figure}[tbph]
\caption{Computational time with respect to the size $n$}
\label{fig:efficiency2}
\centering
\figureskip
\includegraphics[width = \figurewidth, height = \figureheight]{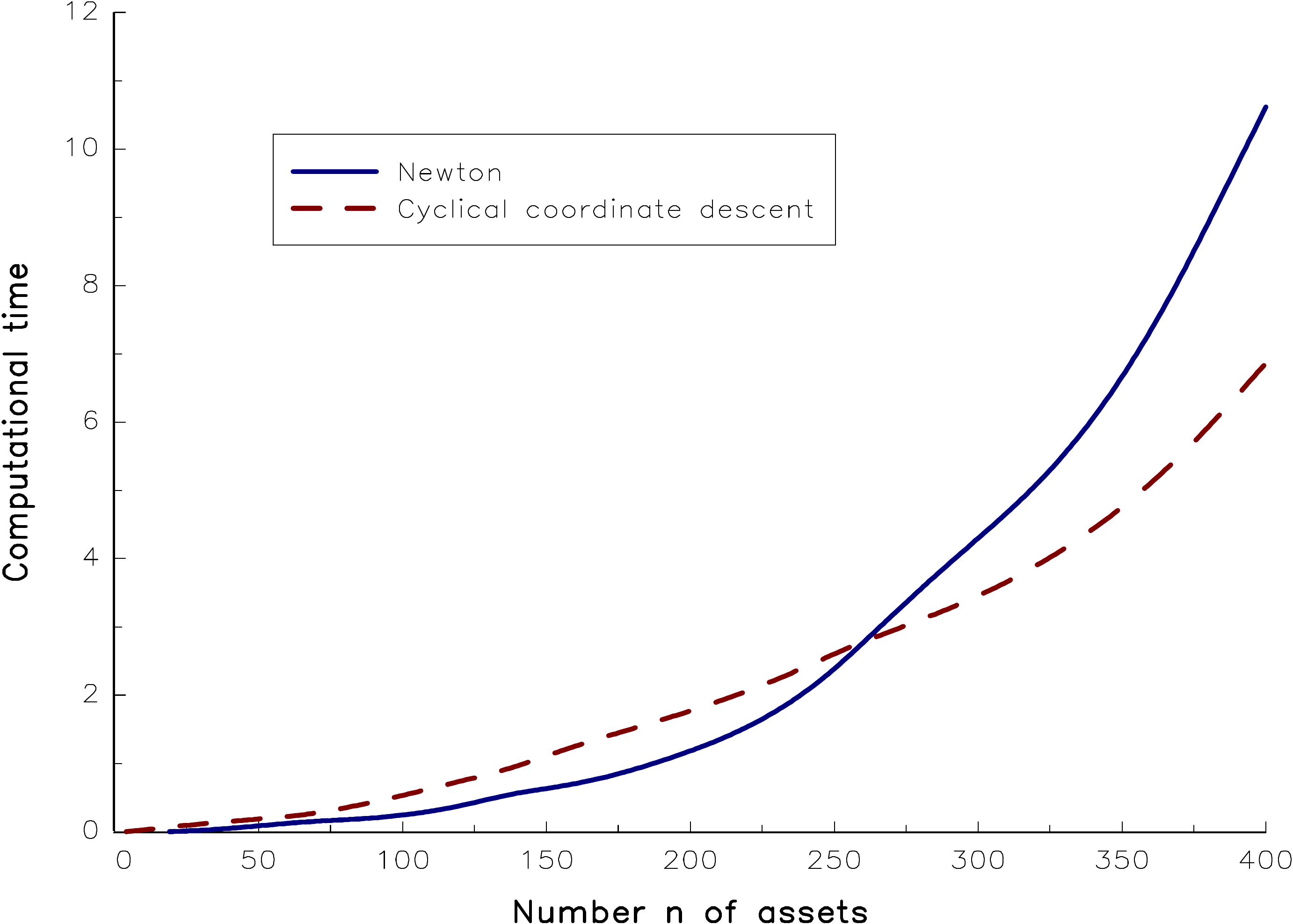}
\end{figure}

We now consider a second application. Using the algorithm of
Davies and Higham (2000), we simulate correlation matrices
such that the singular values are arithmetically distributed.
The computation time of the ERC portfolio
is reported in Table \ref{tab:efficieny1} for different values of the size $n$.
We notice that the Jacobi algorithm does not converge. We also observe
how the computational time of the Newton (or Nesterov) algorithm evolves with respect
to the size of the universe due to the linear system equation.
It follows that the CCD algorithm is more efficient than the Newton algorithm
when the number of assets is larger than $250$ (see Figure \ref{fig:efficiency2}).

\begin{table}[tbph]
\centering
\caption{Computational time with simulated correlation matrix}
\label{tab:efficieny1}
\tableskip
\begin{tabular}{|c|cccc|}
\hline
          $n$  & Jacobi &  Newton       &  Nesterov     & CCD            \\
\hline
  ${\bP}500$   &   NC   &  ${\bP}24$ & ${\bPP}37$ &  ${\bP}13$ \\
      $1000$   &   NC   &      $215$ & ${\bP}384$ &  ${\bP}45$ \\
      $1500$   &   NC   &      $790$ &     $1575$ &      $110$ \\
\hline
\end{tabular}
\end{table}

\begin{remark}
The previous numerical results are sensitive to the programming language and the algorithm
used to compute the Newton step: $\Delta x_{k}=\left[ \partial _{x}^{2}f\left( x_{k}\right)
\right] ^{-1}\partial _{x}f\left( x_{k}\right) $. For instance, it is better to solve the linear system
$\partial _{x}^{2}f\left( x_{k}\right) \Delta x_{k}= \partial _{x}f\left( x_{k}\right) $ using the
Cholesky decomposition than computing the inverse of the symmetric positive definite matrix $\partial _{x}^{2}f\left( x_{k}\right) $%
\footnote{If we use the inverse matrix, the computational time of the Newton algorithm is $35.40$ seconds
instead of $7.90$ seconds for the last example with $n = 1500$.}.
Moreover, the computational time of the CCD algorithm depends on the efficiency of the programming language in terms of loops.
For instance, the Newton algorithm is faster than the CCD algorithm in Matlab and in R because of their poor implementation
of loops. On the contrary, the CCD algorithm is faster than the Newton algorithm in native programming languages
(C or Fortran) and in Gauss.
\end{remark}

\appendix

\section{Existing algorithms}
\label{appendix:algorithm}

In what follows, we consider the standard risk parity approach when
the risk measure $\mathcal{R}\left( x\right) $ is the portfolio
volatility $\sigma \left( x\right) $. The original algorithms were
developed to compute the ERC portfolio, but the extension to the RB
portfolio is straightforward.

\subsection{The SQP algorithm}

Maillard \textsl{et al.} (2010) propose to compute the RB portfolio
by considering the following optimization problem:
\begin{eqnarray*}
x^{\star } &=&\arg \min \sum_{i=1}^{n}\sum_{j=1}^{n}\left( \frac{\mathcal{RC}%
_{i}\left( x\right) }{b_{i}}-\frac{\mathcal{RC}_{j}\left( x\right) }{b_{j}}%
\right) ^{2} \\
&\text{u.c.}&\mathbf{1}^{\top }x=1\quad \text{and}\quad
\mathbf{0}\leq x\leq \mathbf{1}
\end{eqnarray*}
This convex problem can be solved using the sequential quadratic
programming (or SQP) algorithm. However, this method may be improved
by considering a slight modification of the objective function:
\begin{eqnarray*}
x^{\star } &=&\arg \min \sum_{i=1}^{n}\left( \frac{x_{i}\left(
\Sigma
x\right) _{i}}{\sigma ^{2}\left( x\right) }-b_i\right) ^{2} \\
&\text{u.c.}&\mathbf{1}^{\top }x=1\quad \text{and}\quad
\mathbf{0}\leq x\leq \mathbf{1}
\end{eqnarray*}
In this case, we can compute analytically the associated gradient
and Hessian matrices in order to accelerate the computation time.

\subsection{The Jacobi algorithm}

Let $\beta _{i}\left( x\right) $ be the beta of asset $i$ with
respect to the portfolio $x$. We have:
\begin{equation*}
\beta _{i}\left( x\right) =\frac{\left( \Sigma x\right) _{i}}{\sigma
^{2}\left( x\right) }
\end{equation*}
In the RB portfolio, the amounts of beta are proportional to the
risk budgets:
\begin{equation*}
x_{i}\beta _{i}\left( x\right) \propto b_{i}
\end{equation*}
Chaves \textsl{et al.} (2012) suggest to use the Jacobi power method
to find the fixed point. It consists in iterating the previous
formula:
\begin{equation*}
x_{i,k+1}=\frac{b_{i}/\beta _{i}\left( x_{k}\right) }{\sum_{j=1}^{n}b_{j}/%
\beta _{j}\left( x_{k}\right) }
\end{equation*}
where $k$ is the iteration index. Here, the $\beta _{i}\left(
x_{k}\right) $ values are calculated with respect to the portfolio
$x_{k}$ and are used to compute the new weights $x_{k+1}$.

\subsection{The Newton-Nesterov algorithm}

Chaves \textsl{et al.} (2012) propose to apply the Newton method to
the problem (\ref{eq:rb1}). However, the algorithm may have some difficulties to
converge especially when the number of assets is large and the discrepancy between correlations
is large. Spinu (2013)
notices that the associated Lagrange function (\ref{eq:rb2}) is self-concordant and
suggests to use the theory developed by Nesterov (2004) to improve
the efficiency of the algorithm.

\subsubsection{The Newton\ algorithm with self-concordant functions}

Nesterov (2004) consider the following optimization problem:
\begin{equation*}
x^{\ast }=\underset{x\in \limfunc{dom}f}{\arg \min }f\left( x\right)
\end{equation*}
when $f\left( x\right) $ is self-concordant\footnote{%
Let $\phi \left( x;t\right) =f\left( x+tu\right) $ where $x\in
\limfunc{dom}f $ and $u\in \mathbb{R}^{n}$. The function is
self-concordant if it verifies the technical property:
\begin{equation*}
\left| D^{3}f\left( x\right) \left[ u,u,u\right] \right| \leq
M_{f}\left\| u\right\| _{f^{\prime \prime }\left( x\right) }^{3/2}
\end{equation*}
where $D^{3}f\left( x\right) \left[ u,u,u\right] =\partial
_{t}^{3}\phi \left( x;t\right) $, $\left\| u\right\| _{f^{\prime
\prime }\left( x\right) }=\sqrt{u^{\top }f^{\prime \prime }\left(
x\right) u}$ and $M_{f}$ is a positive constant. The underlying
idea of this technical property is to define objective functions for
which there is no convergence problem.}. Let us define $\lambda
_{f}\left( x\right) $ as follows:
\begin{equation*}
\lambda _{f}\left( x\right) =\sqrt{\partial _{x}f\left( x\right)
^{\top } \left[ \partial _{x}^{2}f\left( x\right) \right]
^{-1}\partial _{x}f\left( x\right) }
\end{equation*}
Nesterov (2004) shows that the solution of the problem exists and is
unique if $\lambda _{f}\left( x\right) <1$ and the Hessian is not
degenerate. Moreover, he derives the region of the quadratic
convergence for the Newton algorithm, which is defined as follows:
$\lambda _{f}\left( x\right) <\lambda ^{\ast }$ where $\lambda
^{\ast
}=\left( 3-\sqrt{5}\right) /2$. In this case, one can guarantee that $%
\lambda _{f}\left( x_{k+1}\right) <\lambda _{f}\left( x_{k}\right) $ where $%
x_{k}$ is the Newton solution at the iteration $k$. Finally, the
Newton algorithm applied to self-concordant functions becomes:

\begin{enumerate}
\item  Damped phase\newline
While $\lambda _{f}\left( x_{k}\right) \geq \beta $ where $\beta \in
\left[ 0,\lambda ^{\ast }\right] $, we apply the following
iteration:
\begin{equation*}
x_{k+1}=x_{k}-\frac{1}{1+\lambda _{f}\left( x_{k}\right) }\Delta
x_{k}
\end{equation*}
where $\Delta x_{k}=\left[ \partial _{x}^{2}f\left( x_{k}\right)
\right] ^{-1}\partial _{x}f\left( x_{k}\right) $.

\item  Quadratic phase\newline
When $\lambda _{f}\left( x_{k}\right) <\beta $, we apply the
standard Newton iteration:
\begin{equation*}
x_{k+1}=x_{k}-\Delta x_{k}
\end{equation*}
\end{enumerate}

\subsubsection{Application to the risk parity problem}

Spinu (2013) apply the previous algorithm to the Lagrange
function%
\footnote{We can set the Lagrange coefficient $\lambda$ equal to 1
because of the scaling property of the RB portfolio (Roncalli,
2013).} :
\begin{equation}
f\left(y\right) = \frac{1}{2}y^{\top} C y - \sum_{i=1}^{n} b_i \ln
y_i
\end{equation}
where $C$ is the correlation matrix of asset returns.
He deduces that the gradient and Hessian matrices are:
\begin{eqnarray*}
\partial _{x}f\left( y_{k}\right)  &=&Cy_{k}-by_{k}^{-1} \\
\partial _{x}^{2}f\left( y_{k}\right)  &=&C+\limfunc{diag}\left(
by_{k}^{-2}\right)
\end{eqnarray*}
Moreover, Spinu (2013) proposes to replace $\lambda _{f}\left(
y_{k}\right) $ by $\delta _{f}\left( y_{k}\right) =\left\| \Delta
y_{k}/y_{k}\right\| _{\infty }$ in
the Newton algorithm in order to reduce the computation time\footnote{%
Spinu (2013) takes $\beta =0.95\times \lambda ^{\ast }$.}. For the
initial
point $x_{0}$, he considers the scaled equally-weighted portfolio $%
x_{0}=\left( \mathbf{1}^{\top }C\mathbf{1}\right) ^{-1/2}\cdot
\mathbf{1}$. Finally, we deduce the RB portfolio by rescaling the
solution $y^{\star }$ by the volatilities:
\begin{equation*}
x_{i}^{\star }=\frac{\sigma _{i}^{-1}y_{i}^{\star
}}{\sum_{j=1}^{n}\sigma _{j}^{-1}y_{j}^{\star }}
\end{equation*}
The method proposed by Spinu (2013) improves the Newton
method described in Chaves \textsl{et al.} (2012)
when it fails to converge.

\end{document}